\documentclass[aps,prl,reprint,superscriptaddress,showpacs,floatfix,longbibliography]{revtex4-1}

\usepackage{graphicx}
\usepackage{amssymb,amsmath}
\usepackage{dcolumn}
\usepackage{bm}
\usepackage[mathlines]{lineno}
\usepackage{color}
\usepackage{soul}
\usepackage{amssymb,amsmath,amsthm}

\begin{document}

\title{Continuum approach for
       long-wavelength acoustic phonons in quasi-2D structures}

\author{Dan~Liu}
\affiliation{Physics and Astronomy Department,
             Michigan State University,
             East Lansing, Michigan 48824, USA}

\author{Arthur~G.~Every}
\affiliation{School of Physics,
             University of the Witwatersrand,
             Private Bag 3,
             2050 Johannesburg,
             South Africa
}

\author{David Tom\'{a}nek}
\email
            {tomanek@pa.msu.edu}%
\affiliation{Physics and Astronomy Department,
             Michigan State University,
             East Lansing, Michigan 48824, USA}

\date{\today} 

\begin{abstract}
As an alternative to atomistic calculations of long-wavelength
acoustic modes of atomically thin layers, which are known to
converge very slowly, we propose a quantitatively predictive and
physically intuitive approach based on continuum elasticity
theory. We describe a layer, independent of its thickness, by a
membrane, characterize its elastic behavior by a $(3{\times}3)$
elastic matrix as well as the flexural rigidity. We present simple
quantitative expressions for frequencies of long-wavelength
acoustic modes, which we determine using 2D elastic constants
calculated by {\em ab initio} Density Functional Theory. The
calculated spectra accurately reproduce observed and calculated
long-wavelength phonon spectra of graphene and phosphorene, the
monolayer of black phosphorus. Our approach also correctly
describes the observed dependence of the radial breathing mode
frequency on the diameter of carbon fullerenes and nanotubes.
\end{abstract}



\maketitle

\section{Introduction}

\begin{figure}[b]
\includegraphics[width=1.0\columnwidth]{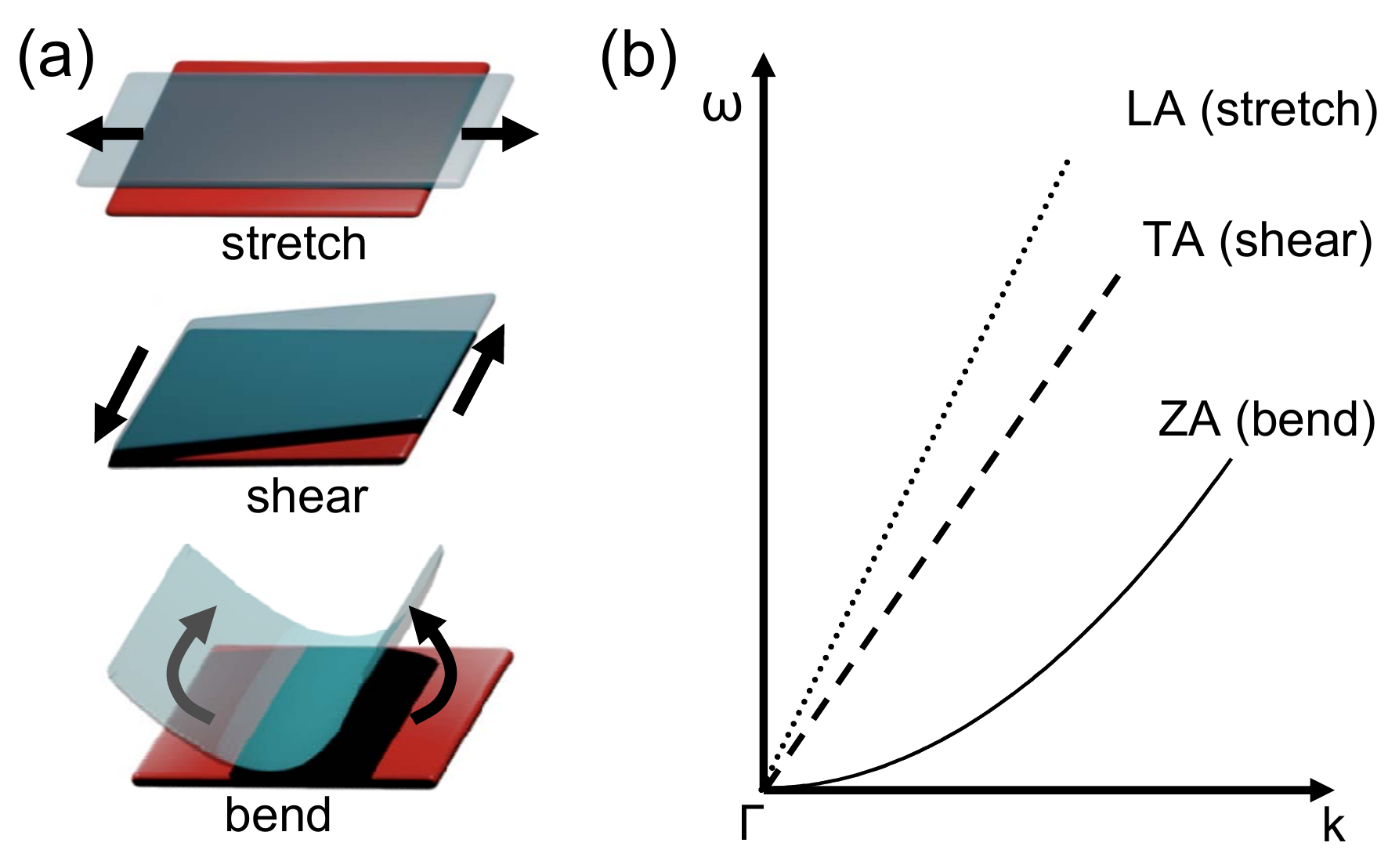}
\caption{(Color online) (a) Schematic representation of possible
distortions of an elastic membrane. (b) Schematic dependence of
the longitudinal acoustic (LA), in-plane transverse acoustic (TA)
and the flexural acoustic (ZA) mode vibration frequencies on the
crystal momentum near the center of the Brillouin zone.
\label{fig1} }
\end{figure}

With the emergence of graphene as one of the hottest research
topics in recent years, interest in quasi-2D materials has been
rising steadily. An important characteristic of these systems are
phonon spectra. State-of-the-art atomistic calculations of phonon
frequencies based on {\em ab initio} Density Functional Theory
(DFT) start with the calculation of the Hessian (or
force-constant) matrix using energy differences associated with
finite displacements of all atoms in the unit cell. The phonon
frequencies are then obtained from the eigenvalues of the
dynamical matrix constructed from the Hessian matrix. This
approach works very well for all phonons except for
long-wavelength acoustic modes in quasi-2D systems. There, all
practical implementations require the use of very large
supercells, dense k-meshes, and a highly converged basis to obtain
a sufficiently accurate dynamical matrix. Even small inaccuracies
caused by computer resource limitations commonly lead to imaginary
frequencies in long-wavelength flexural ZA
modes~\cite{{Peng16},{Yu16}}. This artifact has nothing to do with
a structural instability, but is rather intrinsic to the way the
dynamical matrix is constructed and diagonalized. Even though this
shortcoming does not affect other modes much, its unphysical
consequences have been discussed widely. So far, no practicable,
predictive and accurate alternative approach has been proposed to
determine the frequency of long-wavelength ZA modes, which --
among others -- play an important role in thermal conductivity of
graphene nanoribbons~\cite{flex-graphene16}.

To address the long-standing problem with the computation of
long-wavelength flexural ZA modes in atomically thin layers, we
propose a quantitatively predictive and physically intuitive
approach based on continuum elasticity theory. We describe a
layer, independent of its thickness, by a membrane, characterize
its elastic behavior by a $(3{\times}3)$ elastic matrix as well as
the flexural rigidity $D$. We present simple quantitative
expressions for frequencies of long-wavelength acoustic ZA and --
for the sake of completeness -- also for the longitudinal acoustic
(LA) and transverse acoustic (TA) modes, which we determine using
2D elastic constants calculated by {\em ab initio} Density
Functional Theory. The calculated spectra accurately reproduce
observed and calculated long-wavelength phonon spectra of graphene
and phosphorene, the monolayer of black phosphorus. Our approach
also correctly describes the observed dependence of the radial
breathing mode frequency on the diameter of carbon fullerenes and
nanotubes.

Use of continuum elasticity theory as a means to determine the
frequencies of long-wavelength acoustic phonons is well
established~\cite{Love-book}, including its extension to plates of
finite thickness~\cite{{Maznev-Every95},{Malekzadeh14}}. Elastic
constants have also been related to specific phonon modes in
graphene~\cite{{Lu09},{Lambin14},{Arash2014},{RevModPhysCastro}}.
Yet independent of thickness, any plate can be mapped onto a 2D
elastic membrane. Even though quantifying its elastic response to
uniform static stress is all that is needed to correctly reproduce
all three acoustic branches in the long-wavelength limit, this
knowledge has been used only in a limited fashion to calculate
phonon spectra of quasi-2D systems. In the following, we introduce
a $(3{\times}3)$ elastic stiffness matrix for 2D membranes and
relate it to the commonly used $(6{\times}6)$ elastic matrix for
3D systems. We then derive quantitatively predictive, simple
expressions for long-wavelength acoustic phonon frequencies in
these structures based on the 2D elastic constants and flexural
rigidity D, which we determine using static DFT calculations for
the deformation energy of unit cells with few atoms. The quadratic
frequency dependence on the crystal momentum for ZA and the linear
dependence for LA and TA modes is quantitatively reproduced near
the Brillouin zone center using these 2D elastic constants and
$D$. Clearly, our approach is limited to long-wavelength acoustic
modes. To obtain the full phonon spectrum, these results can be
combined with those of atomistic DFT calculations, which do not
display convergence problems for optical and short-wavelength
acoustic modes.

\section{Continuum approach for long-wavelength acoustic modes
         of an elastic membrane}

A free-standing thin slab of any substance, independent of its
effective thickness, can be mapped onto an elastic membrane that
resists deformations, as illustrated in Fig.~\ref{fig1}(a). In the
harmonic limit, the elastic response of this two-dimensional
system, considered to lie in the $x-y$ plane, is described by the
$(3{\times}3)$ 2D elastic stiffness matrix, which is given in
Voigt notation by
\begin{eqnarray}
\left( \begin{array}{ccc}
 c_{11} & c_{12} & 0 \\
 c_{12} & c_{22} & 0 \\
  0     &   0    & c_{66}
       \end{array}
\right) \,.%
\label{Eq1}
\end{eqnarray}
$c_{11}$ and $c_{22}$ describe the longitudinal strain-stress
relationship along the $x-$ and $y-$direction, respectively.
$c_{66}$ describes the elastic response to in-plane shear. In an
elastic isotropic plate, $c_{11}=c_{22}$,
$c_{66}=(c_{11}-c_{12})/2$, and the Poisson ratio
${\alpha}=c_{12}/c_{11}$. In a 2D plate represented by a membrane,
the dimension of the elastic stiffness constants $c_{ij}$ is
(N/m). The elastic response of the corresponding 3D system
consisting of weakly interacting layers separated by interlayer
spacing $d_{il}$ is described in Voigt notation by a
$(6{\times}6)$ $C_{ij}$ matrix of elastic stiffness constants with
the dimension (N/m$^2$). The relation between the two elastic
matrices is given by $c_{ij}=d_{il}{\cdot}C_{ij}$.

The flexural response to out-of-plane stress is described by the
flexural rigidity $D$ of the plate, which may be anisotropic. $D$
can be calculated by considering the energy cost associated with
rolling up a rectangle of length $L$ to a tube with diameter $d$.
Assuming that $L$ and $d$ are large enough to ignore edge effects,
we obtain
\begin{equation}
D = \frac{1}{2}{\epsilon_b} d^2 \,,%
\label{Eq2}
\end{equation}
where ${\epsilon_b}$ is the bending strain energy divided by the
surface area of the tube, which is close to the area of the
initial rectangle.

\section{Calculation of acoustic phonon modes
         of an elastic membrane in the continuum limit}

It is well established that near the of the Brillouin zone center,
the frequency of LA and TA show a linear dependence on the crystal
momentum $k$, whereas the ZA mode frequency increases as $k^2$, as
seen in Fig.~\ref{fig1}(b). As shown in the Appendix, the
frequencies of the three acoustic modes of an elastic membrane can
be determined quantitatively using only the elastic constants
$c_{11}$, $c_{22}$, $c_{66}$ and $D$. These elastic constants can
be either calculated or obtained experimentally. Also needed is
the 2D mass density $\rho_{2D}$, which is easily determined using
the atomic mass numbers and the area of the optimized unit cell.

For acoustic modes with linear dispersion, we get
\begin{equation}
\omega_{LA,1} = \sqrt{\frac{c_{11}}{\rho_{2D}} } ~ k%
\label{Eq3}
\end{equation}
for the longitudinal acoustic (LA) mode along the $x-$direction,
with the value of the square root giving the longitudinal speed of
sound in the $x-$direction. Similarly, we get
\begin{equation}
\omega_{LA,2} = \sqrt{\frac{c_{22}}{\rho_{2D}}} ~ k%
\label{Eq4}
\end{equation}
for the LA mode along the $y-$direction, with the value of the
square root giving the longitudinal speed of sound in the
$y-$direction. Finally, we get
\begin{equation}
\omega_{TA} = \sqrt{\frac{c_{66}}{\rho_{2D}}} ~ k%
\label{Eq5}
\end{equation}
for the in-plane transverse acoustic (TA) modes in the $x-$ and
$y-$ directions, with the value of the square root giving the
transverse speed of sound in those directions. In anisotropic
plates, acoustic modes in an arbitrary in-plane direction have
mixed LA and TA character, and their speed varies in a complicated
way with the direction~\cite{Maznev-Every95}.

As shown in the Appendix, the flexural ZA mode displays an unusual
quadratic frequency dependence on the crystal momentum. Its
frequency is given by
\begin{equation}
\omega_{ZA} = \sqrt{\frac{D}{\rho_{2D}}} ~ k^2 \,,%
\label{Eq6}
\end{equation}
where the value of the flexural rigidity $D$ depends on the
bending direction in an anisotropic material.

The schematic dependence of the acoustic mode frequencies $\omega$
of a zero-thickness plate on the crystal momentum $\bf k$, given
by Eqs.~(\ref{Eq3})-(\ref{Eq6}), is shown in Fig.~\ref{fig1}(b).
We note that these expressions, albeit not in the notation of the
2D elastic matrix, have been obtained
previously~\cite{{RevModPhysCastro},{ChaikinLubensky},{NelsonStatistical}}.

Expressions describing the deformation of an elastic membrane can
also be used the determine the frequency of the radial breathing
mode (RBM). Previously derived theoretical
expressions~\cite{Ando02} and experimental observations are being
used commonly as an indirect way to determine the diameter of
carbon fullerenes and nanotubes~\cite{DT227}.

As shown in the Appendix, we obtain
\begin{equation}
\omega_{C_n} = \frac{2}{d} \sqrt{\frac{2c_{11}}{\rho_{2D}}} \,,%
\label{Eq7}
\end{equation}
for the RBM frequency of a spherical C$_n$ fullerene with diameter
$d$. Similarly, the RBM frequency of a carbon nanotube (CNT) of
radius $d$ is given by
\begin{equation}
\omega_{CNT} = \frac{2}{d} \sqrt{\frac{c_{11}}{\rho_{2D}}} \,.%
\label{Eq8}
\end{equation}

\section{Computational approach to determine the elastic constants}

We determine the elastic response of atomically thin graphene and
phosphorene monolayers using {\em ab initio} density functional
theory (DFT) as implemented in the {\textsc SIESTA}~\cite{SIESTA}
code. We used the Perdew-Burke-Ernzerhof (PBE)~\cite {PBE}
exchange-correlation functional, norm-conserving Troullier-Martins
pseudopotentials~\cite{Troullier91}, and a double-$\zeta$ basis
including polarization orbitals. To determine the energy cost
associated with in-plane distortions, we sampled the Brillouin
zone of a 3D superlattice of non-interacting layers by a
$20{\times}20{\times}1$ $k$-point grid~\cite{Monkhorst-Pack76}. To
determine the strain energy associated with flexural motion, we
constructed and optimized single-wall nanotubes and sampled their
3D superlattice by a $20{\times}1{\times}1$ $k$-point grid. We
used a mesh cutoff energy of $180$~Ry and an energy shift of
10~meV in our self-consistent total energy calculations, which has
provided us with a precision in the total energy of
${\leq}2$~meV/atom.

\section{Results}

\subsection{Graphene}

Graphene is known to display isotropic elastic behavior. Our DFT
calculations yield the $(3{\times}3)$ 2D elastic stiffness matrix
of Eq.~(\ref{Eq1})
\begin{eqnarray}
\left( \begin{array}{ccc}
 c_{11} & c_{12} & 0 \\
 c_{12} & c_{22} & 0 \\
  0     &   0    & c_{66}
       \end{array}
\right) = %
\left( \begin{array}{ccc}
 352.6 & 59.6  & 0 \\
 59.6  & 352.6 & 0 \\
   0   &   0   & 146.5
       \end{array}
\right) N/m\,.%
\label{Eq9}
\end{eqnarray}
These values are in very good agreement with experimental and
other theoretical results~\cite{PhysRevLettZakharchenko},
including the value of the Poisson ratio
${\alpha}=c_{12}/c_{11}=0.17$, which is very close to the observed
value of 0.19 based on HREELS~\cite{Politano12}.

The calculated 2D mass density of graphene is
${\rho_{2D}}=0.743{\cdot}10^{-6}$~kg/m$^2$ 
and the calculated value of the flexural rigidity %
$D = 1.40$~eV$ = 0.224$~GPa${\cdot}$nm$^3$ lies close to the
previously estimated value~\cite{RevModPhysCastro}
$D{\approx}1.0$~eV. Using the numerical values listed in
Eq.~(\ref{Eq9}) and the above value of $D$, we can determine the
three acoustic branches of graphene near the $\Gamma$ point using
Eqs.~(\ref{Eq3})-(\ref{Eq6}). Our results are presented in
Fig.~\ref{fig2}(a), superposed to those of a more recent {\em ab
initio} calculation~\cite{Wirtz2004141} that agrees very well with
with the observed and fitted phonon spectrum of
graphene~\cite{Jishi93}.

First of all, we notice an excellent agreement with the linear LA
and TA modes, which indicate that the calculated speed of sound
agrees with the observation. Specifically, the calculated speed of
sound with in-plane longitudinal polarization,
$v_{LA,th}=22.1$~km/s, agrees very well with the observed value
$v_{LA,expt}=22.0$~km/s obtained using HREELS~\cite{Politano12}.
Similarly, the speed of sound with in-plane transverse
polarization, $v_{TA,th}=14.2$~km/s, agrees very well with the
observed value~\cite{Politano12} of $v_{TA,expt}=14.0$~km/s.

Agreement between calculated and observed ZA modes indicates that
the calculated flexural rigidity value correctly reproduces the
elastic response of graphene to bending.

\subsection{Phosphorene}

Unlike graphene, phosphorene is strongly anisotropic. It is much
softer under compression along the $x-$ (or $\bf a_1-$) direction
than along the $y-$ (or $\bf a_2-$) direction. The optimized
rectangular unit cell is defined by $a_1=4.63$~{\AA} and
$a_2=3.35$~{\AA} according to our DFT studies. With 4 atoms per
unit cell, the 2D mass density of phosphorene is
${\rho_{2D}}=1.34{\cdot}10^{-6}$~kg/m$^2$. Our numerical values
$c_{11}=24.4$~N/m and $c_{22}=94.6$~N/m reflect the strong
anisotropy in the in-plane longitudinal elastic response. The
calculated speed of sound with LA polarization is
$v_{LA_X,th}=4.3$~km/s along the soft ${\Gamma}-X$ direction and
$v_{LA_Y,th}=8.4$~km/s along the stiff ${\Gamma}-Y$ direction.

\begin{figure*}[t]
\includegraphics[width=1.5\columnwidth]{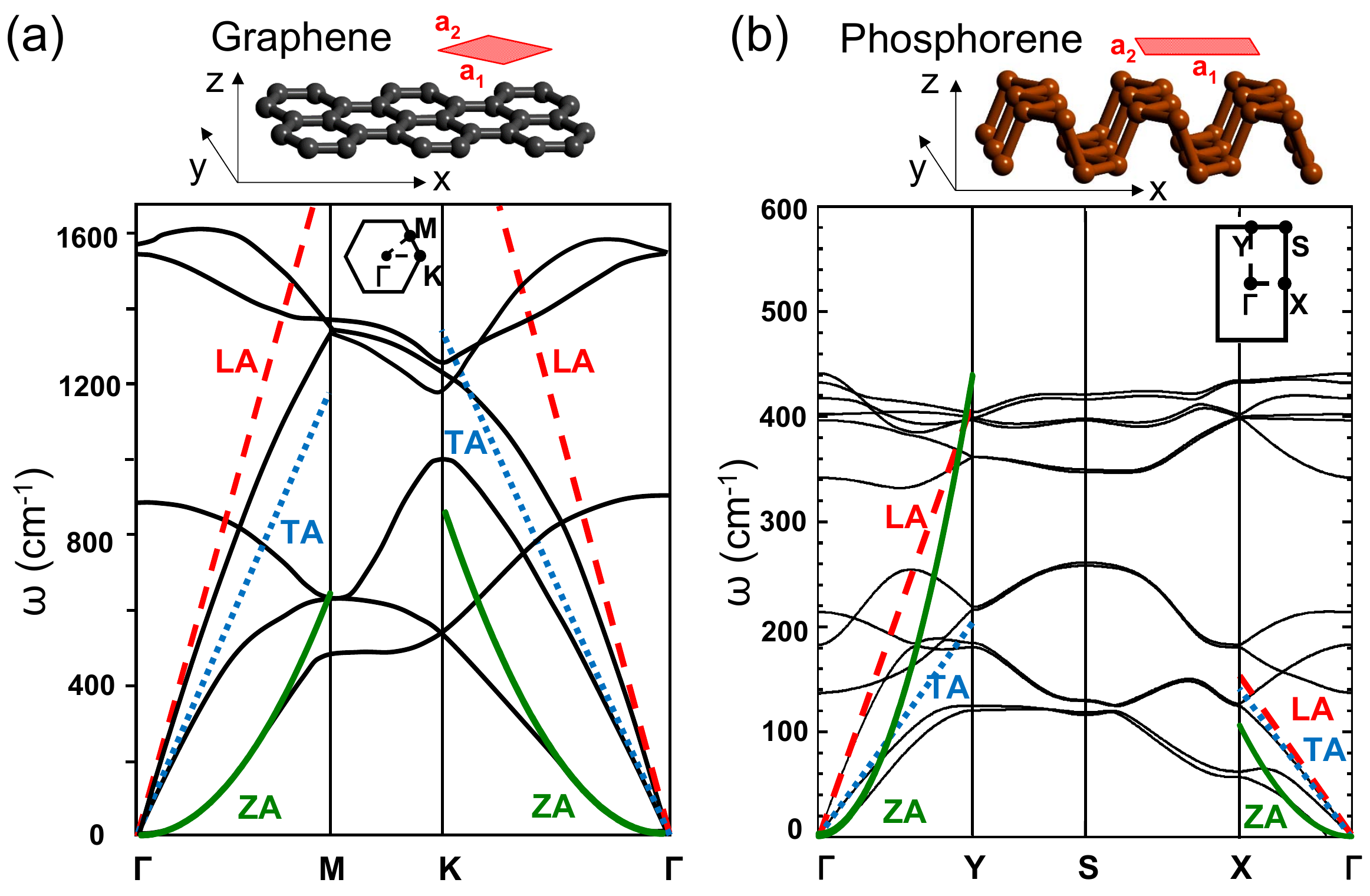}
\caption{(Color online) Phonon spectra of (a) graphene, reproduced
from Reference~\protect\onlinecite{Wirtz2004141} and (b)
phosphorene, a monolayer of black phosphorus, reproduced from
Reference~\protect\onlinecite{DT230}, shown by solid lines.
Superposed to the spectra are continuum results for the three
acoustic phonon modes in different high-symmetry directions,
evaluated near $\Gamma$, with the longitudinal acoustic (LA,
dashed lines), in-plane transverse acoustic (TA, dotted lines),
and the flexural acoustic (ZA, solid lines) modes. Ball-and-stick
models of the structure, including the primitive unit cells, are
shown in the top panels. The Brillouin
zones are shown as insets in the phonon spectra.%
\label{fig2}}
\end{figure*}

The transverse acoustic phonon frequency depends on the in-plane
shear and is described by our calculated value $c_{66}=22.1$~N/m.
The corresponding speed of sound with TA polarization is
$v_{TA,th}=4.1$~km/s.

Finally, we find also the flexural rigidity to be highly
anisotropic. We find
$D({\Gamma}-X)=1.55$~eV$=0.248$~GPa${\cdot}$nm$^3$ when bending
phosphorene along the $x-$ (or $\bf a_1-$) direction, yielding a
tube with its axis aligned along the $y-$ (or $\bf a_2-$)
direction. Bending along the $y-$ (or $\bf a_2-$) direction, we
find $D({\Gamma}-Y)=7.36$~eV$=1.179$~GPa${\cdot}$nm$^3$.

These data are sufficient to reproduce the three acoustic branches
of phosphorene along the ${\Gamma}-X$ and ${\Gamma}-Y$ direction
near $\Gamma$ using Eqs.~(\ref{Eq3})-(\ref{Eq6}) and are presented
in Fig.~\ref{fig2}(b). In analogy to graphene, our results are
superposed to phonon spectra , which -- in absence of experimental
phonon spectra -- are based on atomistic DFT calculations of
phosphorene~\cite{DT230}. We notice a particularly good agreement
between the two approaches along the stiff ${\Gamma}-Y$ direction.
As we expand in the Discussion, the continuum results deviate from
those of the atomistic study along the soft ${\Gamma}-X$
direction, which -- at close inspection -- predicts small
imaginary ZA frequencies near $\Gamma$.

\begin{figure}[b]
\includegraphics[width=1.0\columnwidth]{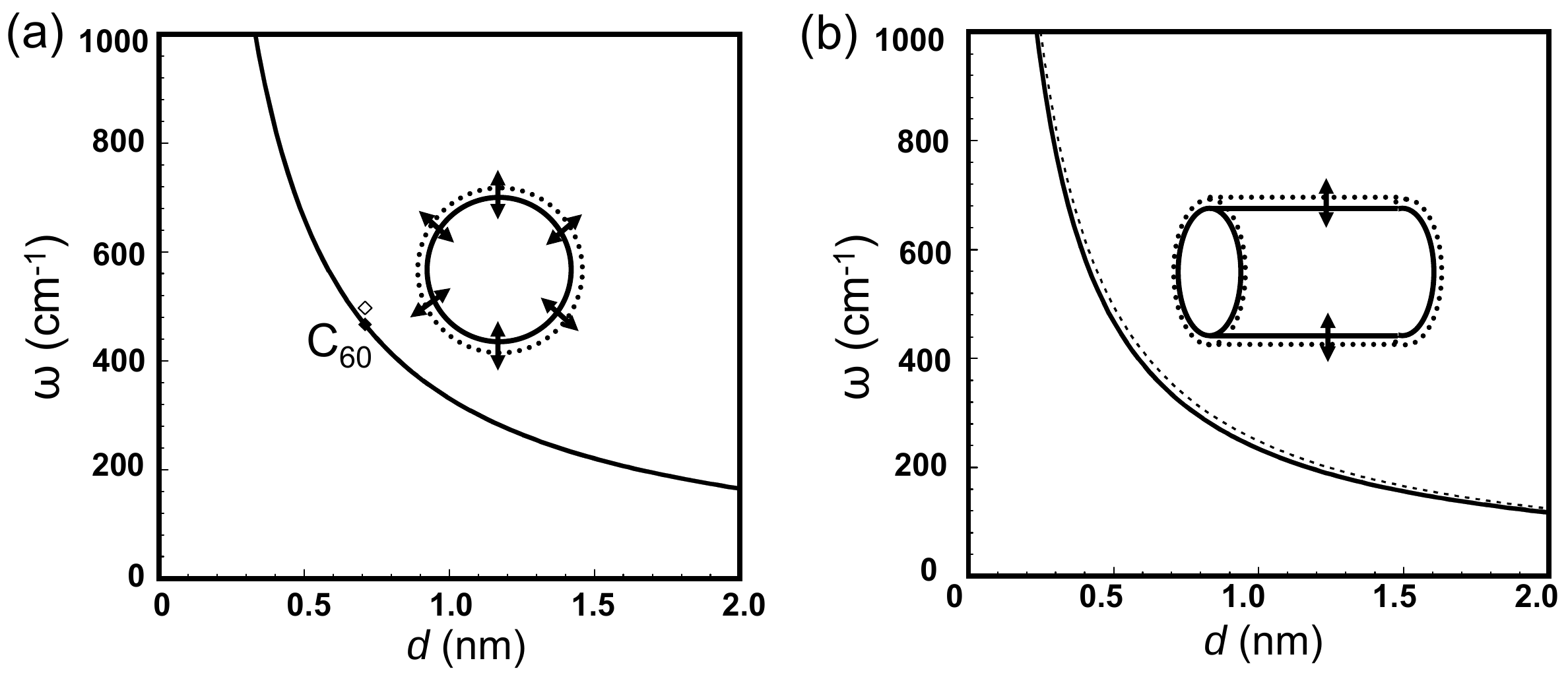}
\caption{Radial breathing mode (RBM) in (a) fullerenes and (b)
carbon nanotubes as a function of their diameter $d$. Our
prediction is shown by the solid line. Experimental and
theoretical RBM frequencies for the only observed fullerene with
spherical symmetry, C$_{60}$, are shown by the data points in (a).
The experimentally well-established
relationship~\protect\cite{{RBM-Rao97},{DT227}}
$\omega=248$~cm$^{-1}{\times}(1$~nm$/d)$ in nanotubes in (b) is
represented by the dashed line and our calculation is
shown by the solid line.%
\label{fig3} }
\end{figure}

\subsection{Vibration spectra of carbon fullerenes and nanotubes}

The radial breathing mode (RBM) frequency $\omega_{RBM}$ of C$_n$
fullerenes may be calculated using Eq.~(\ref{Eq7}) and that of
carbon nanotubes using Eq.~(\ref{Eq8}), in combination with the
elastic constants provided in the section on graphene. The
expected dependence of $\omega_{RBM}$ on the diameter $d$ is
displayed in Fig.~\ref{fig3}(a) for fullerenes and in
Fig.~\ref{fig3}(b) for carbon nanotubes (CNTs). The RBM mode is
Raman active and its frequency is known to depend primarily on the
diameter. Thus, the RBM frequency is commonly used to judge the
diameter of carbon nanostructures.

According to Eq.~(\ref{Eq7}), the RBM frequency of spherical
fullerenes should scale inversely with their diameter. But only
one spherical fullerene, namely C$_{60}$ with $d=7.1$~{\AA}, forms
a molecular solid. We find the predicted value
$\omega_{RBM,th}({\rm C_{60}})=467$~cm$^{-1}$ to lie very close to
the observed RBM frequency $\omega_{RBM,expt}({\rm
C_{60}})=497$~cm$^{-1}$.

The well-documented observed diameter dependence of the RBM in
nanotubes~\cite{DT227}, $\omega_{RBM,expt}({\rm
CNT})=248$~cm$^{-1}{\times}$(1~nm/$d$), is reproduced by the
dashed line in Fig.~\ref{fig3}(b). Based on continuum theory and
Eq.~(\ref{Eq8}), we find $\omega_{RBM,th}({\rm
CNT})=234$~cm$^{-1}{\times}$(1~nm/$d$) in very good agreement with
the observed behavior.

\section{Discussion}

So far, the most common description of a layer by continuum
elasticity theory has been that of a finite-thickness plate
consisting of a material characterized by a $(6{\times}6)$ elastic
stiffness matrix. As we show here, this approach is unnecessarily
cumbersome, since every 2D system of finite thickness may be
mapped onto an elastic membrane with the same mass density. The
resistance of the realistic system, such as graphene or
phosphorene, to stretching, shear and bending becomes that of the
elastic membrane. The advantage of this approach, which does not
suffer from ambiguities about the ``real'' thickness of an atomic
layer, has been discussed before~\cite{{Lu09},{Lambin14}}. The
unconventional units of the 2D elastic stiffness matrix in
Eq.~(\ref{Eq1}) are well adapted to ultra-thin layers.

Our approach appears particularly suitable when describing the ZA
mode in soft, atomically thin atomic layers such as phosphorene.
As mentioned earlier during the discussion of Fig.~\ref{fig2}(b),
the required precision of the dynamical matrix has not been
reached in the calculation of phonon modes near $Gamma$ along the
${\Gamma}-X$ direction~\cite{DT230}. This is a common shortcoming
of {\em ab initio} phonon calculations in layered solids. Among
the eigenvalues of the dynamical matrix, which are proportional to
${\omega}^2$, the one associated with the ZA mode often turns out
to be negative, yielding nominally an imaginary frequency, as a
numerical artifact. The small error in the ZA frequency eigenvalue
is also reflected in other close-lying eigenvalues, such as those
of the LA and TA modes, at the same crystal momentum. This is
clearly reflected in Fig.~\ref{fig2}(b). We believe that the
present continuum approach is much better adapted to describe
long-wavelength acoustic modes and should be preferred to
atomistic phonon calculations near $\Gamma$.

As seen in Fig.~\ref{fig2}(b), the LA and TA mode frequencies are
very similar along the soft ${\Gamma}-X$ direction in phosphorene.
This is unusual, but not unexpected in view of the accordion-like
structure depicted in the ball-and-stick model in the top panel of
Fig.~\ref{fig2}(b). Making an analogy to a real accordion, it
appears equally easy to produce a longitudinal and a transversal
motion while the instrument is being played. Since the TA speed of
sound is the same along the ${\Gamma}-X$ and the ${\Gamma}-Y$, the
TA mode is clearly distinguished in its softness from the LA mode
along the hard ${\Gamma}-Y$ direction. In this rigid direction in
space, small imprecisions in the dynamical matrix play a much less
important role than along the soft direction. Therefore, our
continuum results agree well with the phonon frequencies obtained
using the atomistic approach for all three acoustic branches.

In the Section on the radial breathing motion of carbon
nanostructures, we have shown in Eq.~(\ref{EqA52}) that the RBM
frequency of nanotubes does not show a pure $1/d$ behavior as a
function of the nanotube diameter. For single-wall carbon
nanotubes with typical diameters between $1-2$~nm, the value of
the correction $4D/(c_{11} d^2)$ is indeed negligibly small in
comparison to 1. This need not be the case in nanotubes of other
substances with large values of $D$ and small values of $c_{11}$.
In postulated phosphorene nanotubes~\cite{DT235}, $c_{11}$ differs
significantly from $c_{22}$, and $D$ is strongly anisotropic.

Finally, we have shown in Eq.~(\ref{Eq2}) how to estimate the
value of $D$ in a layered material by calculating the strain
energy in nanotubes with a very large diameter $d$. Optimizing
wide nanotubes using {\em ab initio} techniques is non-trivial.
The values for $D$ quoted in the Section on phosphorene required
DFT calculations containing more than 14 unit cells along the
perimeter of phosphorene nanotubes bent along the soft $x-$ (or
${\bf a_1}-$) direction. For nanotubes bent along the rigid $y-$
(or ${\bf a_2}-$) direction, $D(d{\rightarrow}\infty$ was
extrapolated using 10, 12, 14, 16 and 18 unit cells along the
perimeter of the corresponding phosphorene nanotubes.

\section{Summary and Conclusions}

In conclusion, as a viable alternative to atomistic calculations
of long-wavelength acoustic modes of atomically thin layers, which
are known to converge very slowly, we have proposed a
quantitatively predictive and physically intuitive approach based
on continuum elasticity theory. We describe a layer, independent
of its thickness, by a membrane, and characterize its elastic
behavior by a $(3{\times}3)$ elastic matrix as well as the
flexural rigidity $D$. We have derived simple quantitative
expressions for frequencies of long-wavelength acoustic ZA and --
for the sake of completeness -- also for the longitudinal acoustic
(LA) and transverse acoustic (TA) modes. These frequencies are
determined using 2D elastic constants obtained from {\em ab
initio} DFT calculations for the deformation energy of unit cells
with few atoms. We found that the calculated spectra accurately
reproduce observed and calculated long-wavelength phonon spectra
of graphene and phosphorene, the monolayer of black phosphorus.
Our approach also correctly describes the observed dependence of
the radial breathing mode frequency on the diameter of carbon
fullerenes such as C$_{60}$ and carbon nanotubes.

\section{Appendix}
\setcounter{equation}{0}
\renewcommand{\theequation}{A\arabic{equation}}

The most fundamental way to determine vibration motion in any
system starts with a Lagrangian, from which the Euler-Lagrange
equations of motion can be derived by applying Hamilton's
principle of least action. In the following, we determine the
Lagrangians that describe stretching, shearing and bending of an
infinitely thin elastic plate. Using these Lagrange functions, we
derive the equations of stretching, shearing and bending motion of
the plate. Finally, we derive the Lagrangian describing the radial
breathing motion of carbon fullerenes and nanotubes and determine
the corresponding equations of motion for the RBM in these
nanostructures.

\subsection{Lagrange function of an
            infinitely thin elastic plate under strain}

\subsubsection{Stretching}

Let us consider a thin plate suspended in the $xy-$plane and its
response to in-plane tensile strain applied uniformly along the
$x-$direction,
\begin{equation}
\frac{du_x}{dx} = {\rm const.}%
\label{EqA1}
\end{equation}
The resulting energy density caused by tensile strain along the
$x-$direction is then given by
\begin{equation}
U = \frac{1}{2} c_{11} \left( \frac{du_x}{dx}\right)^2 \,,%
\label{EqA2}
\end{equation}
where $c_{11}$ describes the elastic stiffness to tensile strain.
In the harmonic regime, we will consider only small strain values.
Releasing the strain will cause a vibration in the $x$-direction
with the velocity $v_x=du_x/dt$. Then, the kinetic energy density
will be given by
\begin{equation}
T = \frac{1}{2} {\rho_{2D}} \left( \frac{du_x}{dt}\right)^2 \,,%
\label{EqA3}
\end{equation}
where ${\rho_{2D}}$ is the 2D mass density. The Lagrangian density
is then given by
\begin{eqnarray}
\label{EqA4}
\mathcal{L}\left(\frac{du_x}{dx},\frac{du_x}{dt},x,t\right)
&=& T - U \\
&=& \frac{1}{2} \left[%
        {\rho_{2D}} \left( \frac{du_x}{dt}\right)^2%
      - c_{11}  \left( \frac{du_x}{dx}\right)^2%
            \right] \,. \nonumber%
\end{eqnarray}
In an anisotropic plate, the $x-$ and $y-$directions are not
equivalent. To describe the $y-$response to in-plane tensile
strain applied uniformly along the $y-$direction, we have to
replace $x$ by $y$ and $c_{11}$ by $c_{22}$ in Eqs.~({A1})-({A4}),
yielding the Lagrangian density
\begin{eqnarray}
\label{EqA5}
\mathcal{L}\left(\frac{du_y}{dy},\frac{du_y}{dt},y,t\right)
&=& T - U \\
&=& \frac{1}{2} \left[%
        {\rho_{2D}} \left( \frac{du_y}{dt}\right)^2%
          - c_{22}  \left( \frac{du_y}{dy}\right)^2%
            \right] \,. \nonumber%
\end{eqnarray}

\subsubsection{Shearing}

The derivation of the Euler-Lagrange equation for the shear motion
is very similar to that for the stretching motion. The main
difference is that the displacement $u$ is normal to the
propagation direction. To obtain the corresponding equations, we
need to replace $u_x$ by $u_y$ and $c_{11}$ by $c_{66}$ in
Eqs.~({A1})-({A4}). The Lagrangian density is then given by
\begin{eqnarray}
\label{EqA6}
\mathcal{L}\left(\frac{du_y}{dx},\frac{du_y}{dt},x,t\right)
&=& T - U \\
&=& \frac{1}{2} \left[%
        {\rho_{2D}} \left( \frac{du_y}{dt}\right)^2%
          - c_{66}  \left( \frac{du_y}{dx}\right)^2%
            \right] \,. \nonumber%
\end{eqnarray}

\subsubsection{Bending}

Bending a thin plate suspended in the $xy$-plane plane in order to
to achieve a radius of curvature $R$ requires displacements
$u_z(x)$ along the normal $z$-direction that are described by
\begin{equation}
\frac{d^2u_z}{dx^2} = \frac{1}{R} \,.%
\label{EqA7}
\end{equation}
The resulting bending energy density is then given by
\begin{equation}
U = \frac{1}{2} {D} \left( \frac{d^2u_z}{dx^2}\right)^2 \,,%
\label{EqA8}
\end{equation}
where ${D}$ is the flexural rigidity. In the harmonic regime, we
will consider only small strain values corresponding to
$R{\rightarrow}\infty$. Releasing the strain will cause a
vibration in the $z$-direction with the velocity $v_z=du_z/dt$.
Then, the kinetic energy density will be given by
\begin{equation}
T = \frac{1}{2} {\rho_{2D}} \left( \frac{du_z}{dt}\right)^2 \,,%
\label{EqA9}
\end{equation}
where ${\rho_{2D}}$ is the 2D mass density. This leads to the
Lagrangian density
\begin{eqnarray}
\label{EqA10}
\mathcal{L}\left(\frac{d^2u_z}{dx^2},\frac{du_z}{dt},x,t\right)
&=& T - U \\ \nonumber %
&=& \frac{1}{2} \left[%
        {\rho_{2D}} \left( \frac{du_z}{dt}\right)^2%
      - {D}  \left( \frac{d^2u_z}{dx^2}\right)^2%
               \right] \,.
\end{eqnarray}
%


\subsection{Derivation of Euler-Lagrange equations of motion
for deformations of an infinitely thin elastic plate using
Hamilton's principle}

\subsubsection{Stretching}

The Lagrangian specified in Eq.~(\ref{EqA4}) has the form
$\mathcal{L}(du_x/dx,du_x/dt,x,t)$. In this case, we have two
independent variables, $x$ and $t$, and can define the action $S$
by
\begin{equation}
S = \int_{t_1}^{t_2} dt \int_{x_1}^{x_2} dx %
\mathcal{L}(\frac{du_x}{dx},\frac{du_x}{dt},x,t) \,.%
\label{EqA11}
\end{equation}
Hamilton's principle of least action yields
\begin{equation}
{\delta}S = {\delta}\int_{t_1}^{t_2} dt \int_{x_1}^{x_2} dx %
\mathcal{L}(\frac{du_x}{dx},\frac{du_x}{dt},x,t) = 0\,.%
\label{EqA12}
\end{equation}
This can be modified to
\begin{eqnarray}
{\delta}S
&=& \int_{t_1}^{t_2}dt\int_{x_1}^{x_2}dx %
    [\mathcal{L}(\frac{du_x}{dx}+%
    {\delta}\frac{du_x}{dx},\frac{du_x}{dt}+%
    {\delta}\frac{du_x}{dt},x,t) \nonumber\ \\
& & -\mathcal{L}(\frac{du_x}{dx},\frac{du_x}{dt},x,t)]\,.
\label{EqA13}
\end{eqnarray}
and
\begin{equation}
{\delta}S=\int_{t_1}^{t_2}dt\int_{x_1}^{x_2}dx \
             \left[\frac{\partial\mathcal{L}}{\partial\frac{du_x}{dx}}
             {\delta}\frac{du_x}{dx}
            +\frac{\partial\mathcal{L}}{\partial\frac{du_x}{dt}}
              {\delta}\frac{du_x}{dt} \right ] \,. \\
\label{EqA14}
\end{equation}
This expression can be reformulated to
\begin{equation}
{\delta}S= \int_{t_1}^{t_2}\!\!\!\!dt\int_{x_1}^{x_2}\!\!\!\!dx%
\left[ \frac{d}{dt}\!\!%
\left (\!\!\frac{\partial\mathcal{L}}{\partial\frac{du_x}{dt}}\!\!\right )%
           +  \frac{d}{dx} \left(\!\!%
\frac{\partial\mathcal{L}}{\partial\frac{du_x}{dx}}\!\!%
                \right)\right]{\delta}u_x \,. %
\label{EqA15}
\end{equation}
For ${\delta}S$ to vanish, the quantity in the square brackets in
Eq.~(\ref{EqA15}) must vanish. This leads to the Euler-Lagrange
equation
\begin{equation}
    \frac{d}{dt} \left(%
        \frac{{\partial}\mathcal{L}}{{\partial}\frac{du_x}{dt}}
                \right)%
 +  \frac{d}{dx} \left(%
        \frac{{\partial}\mathcal{L}}{{\partial}\frac{du_x}{dx}}
                \right)%
    = 0 \,.%
\label{EqA16}
\end{equation}
Inserting the Lagrangian of Eq.~(\ref{EqA4}) in the Euler-Lagrange
Eq.~(\ref{EqA16}) yields the wave equation for longitudinal
in-plane vibrations
\begin{equation}
    {\rho_{2D}}\frac{d^2u_x}{dt^2} - c_{11} \frac{d^2u_x}{dx^2} = 0 \,.%
\label{EqA17}
\end{equation}
This wave equation can be solved using the ansatz
\begin{equation}
    u_x = u_{x,0} e^{i(kx-{\omega}t)}%
\label{EqA18}
\end{equation}
to yield
\begin{equation}
    {\rho_{2D}}{\omega}^2 = c_{11} k^2 \,.%
\label{EqA19}
\end{equation}
This finally translates to the desired form
\begin{equation}
    {\omega} = \sqrt{\frac{c_{11}}{\rho_{2D}}} k \,,%
\label{EqA20}
\end{equation}
which is identical to Eq.~(\ref{Eq3}).

In an anisotropic plate, we need to use the Lagrangian of
Eq.~(\ref{EqA5}) to describe motion along the $y-$direction and
obtain
\begin{equation}
    {\omega} = \sqrt{\frac{c_{22}}{\rho_{2D}}} k \,,%
\label{EqA21}
\end{equation}
which is identical to Eq.~(\ref{Eq4}).

\subsubsection{Shearing}

The Lagrangian $\mathcal{L}(du_y/dx,du_y/dt,x,t)$ in
Eq.~(\ref{EqA6}), which describes the shearing motion, has the
same form as the Lagrangian in Eq.~(\ref{EqA4}). To obtain the
equations for shear motion from those for stretching motion, we
need to replace $u_x$ by $u_y$ and $c_{11}$ by $c_{66}$ in
Eqs.~(\ref{EqA11})-(\ref{EqA20}). Thus, the equation for shear
motion becomes
\begin{equation}
    {\omega} = \sqrt{\frac{c_{66}}{\rho_{2D}}} k \,,%
\label{EqA22}
\end{equation}
which is identical to Eq.~(\ref{Eq5}).

\subsubsection{Bending}

The Lagrangian specified in Eq.~(\ref{EqA10}) has the
unconventional form $\mathcal{L}(d^2u_z/dx^2,du_z/dt,x,t)$. In
this case, we have two independent variables, namely $x$ and $t$,
and can define the action $S$ by
\begin{equation}
S = \int_{t_1}^{t_2} dt \int_{x_1}^{x_2} dx\ %
\mathcal{L}(\frac{d^2u_z}{dx^2},\frac{du_z}{dt},x,t) \,.%
\label{EqA23}
\end{equation}
Hamilton's principle of least action yields
\begin{equation}
{\delta}S = {\delta}\int_{t_1}^{t_2} dt \int_{x_1}^{x_2} dx\ %
\mathcal{L}(\frac{d^2u_z}{dx^2},\frac{du_z}{dt},x,t) = 0\,.%
\label{EqA24}
\end{equation}
This can be modified to
\begin{eqnarray}
{\delta}S &=& \int_{t_1}^{t_2}dt \int_{x_1}^{x_2}dx %
[%
\mathcal{L}(\frac{d^2u_z}{dx^2}+%
{\delta}\frac{d^2u_z}{dx^2},\frac{du_z}{dt}+%
{\delta}\frac{du_z}{dt},x,t) \nonumber \\
& & -\mathcal{L}(\frac{d^2u_z}{dx^2},\frac{du_z}{dt},x,t)%
]%
\,.%
\label{EqA25}
\end{eqnarray}
and
\begin{equation}
{\delta}S = \int_{t_1}^{t_2}dt \int_{x_1}^{x_2}dx \
               \left [ \frac{\partial\mathcal{L}}{\partial\frac{d^2u_z}{dx^2}}
               {\delta}\frac{d^2u_z}{dx^2}
             + \frac{\partial\mathcal{L}}{\partial\frac{du_z}{dt}}
                {\delta}\frac{du_z}{dt} \right ] \,. \\
\label{EqA26}
\end{equation}
Finally, we can rewrite this expression as
\begin{equation}
{\delta}S = \int_{t_1}^{t_2}\!\!\!\!dt \int_{x_1}^{x_2}\!\!\!\!dx%
\left [ \frac{d}{dt}\!\!%
\left (\!\!\frac{\partial\mathcal{L}}{\partial\frac{du_z}{dt}}\!\!\right )%
             -  \frac{d^2}{dx^2} \left(\!\!%
\frac{\partial\mathcal{L}}{\partial\frac{d^2u_z}{dx^2}}\!\!%
\right ) \right ]{\delta}u_z \,.%
\label{EqA27}
\end{equation}
For ${\delta}S$ to vanish, the quantity in the square brackets in
Eq.~(\ref{EqA27}) must vanish. This leads to the Euler-Lagrange
equation
\begin{equation}
    \frac{d}{dt} \left(%
        \frac{{\partial}\mathcal{L}}{{\partial}\frac{du_z}{dt}}
                \right)%
 -  \frac{d^2}{dx^2} \left(%
        \frac{{\partial}\mathcal{L}}{{\partial}\frac{d^2u_z}{dx^2}}
                \right)%
    = 0 \,.
\label{EqA28}
\end{equation}
Inserting the Lagrangian of Eq.~(\ref{EqA10}) for flexural motion
in the Euler-Lagrange Eq.~(\ref{EqA28}) yields the wave equation
for flexural vibrations
\begin{equation}
    {\rho_{2D}}\frac{d^2u_z}{dt^2} + {D} \frac{d^4u_z}{dx^4} = 0 \,.
\label{EqA29}
\end{equation}
This wave equation can be solved using the ansatz
\begin{equation}
    u_z = u_{z,0} e^{i(kx-{\omega}t)} %
\label{EqA30}
\end{equation}
to yield
\begin{equation}
    {\rho_{2D}}{\omega}^2 = {D} k^4 \,.%
\label{EqA31}
\end{equation}
This finally translates to the desired form
\begin{equation}
    {\omega} = \sqrt{\frac{{D}}{\rho_{2D}}} k^2 \,,%
\label{EqA32}
\end{equation}
which is identical to Eq.~(\ref{Eq6}).


\subsection{Radial breathing mode of spherical fullerenes}

Let us consider a spherical fullerene molecule with the
equilibrium radius $R$, such as C$_{20}$, C$_{60}$, ... Except for
the presence of 12 pentagons, the surface is covered by hexagonal
carbon rings, so that the 2D mass density ${\rho_{2D}}$ can be
taken as that of graphene. Similarly, $R$ can be estimated using
the number of C atoms in the fullerene and the unit cell area in
graphene. According to continuum elasticity theory, the total
bending strain energy of any such spherical fullerene, independent
of $R$, is given by~\cite{DT071}
\begin{equation}
U_b = 4 \pi D (1 + \alpha) \,,%
\label{EqA33}
\end{equation}
where $D$ is the flexural rigidity and $\alpha$ is the Poisson
ratio of graphene. In other words, changing the radius of the
fullerene by ${\delta}R$ will not affect the total bending energy.

Allowing the equilibrium radius $R$ of the fullerene to change by
${\delta}R$ results, on the other hand, in a quadratic increase of
the tensile strain energy
\begin{equation}
U_t = 2 {\times} \frac{1}{2} \left[ 4 \pi R^2 c_{11}\right]%
      \left( \frac{{\delta}R}{R} \right)^2 \,.%
\label{EqA34}
\end{equation}
The (radial) kinetic energy of a radially expanding or contracting
fullerene, shown schematically in the inset of Fig.~\ref{fig3}(a),
is given by
\begin{equation}
T = \frac{1}{2} \left[ 4 \pi R^2 {\rho_{2D}} \right] %
    \left( \frac{d}{dt}{\delta}R \right)^2
\label{EqA35}
\end{equation}
and the Lagrangian by
\begin{eqnarray}
\label{EqA36}%
\mathcal{L} &=& T - U = T - U_t \\%
 &=& \frac{1}{2} \left[ 4 \pi R^2 {\rho_{2D}} \right] %
     \left( \frac{d}{dt}{\delta}R \right)^2 %
   -  \left[ 4 \pi R^2 c_{11} \right] %
      \left( \frac{{\delta}R}{R} \right)^2 \nonumber \\
 &=& 4 \pi R^2 \left[ \frac{1}{2} {\rho_{2D}}
             \left( \frac{d}{dt}{\delta}R \right)^2 %
           - c_{11} \left( \frac{{\delta}R}{R} \right)^2 %
               \right] \,. \nonumber
\end{eqnarray}
Hamilton's principle of least action
\begin{equation}
  \frac{\partial\mathcal{L}}{\partial({\delta}R)}
- \frac{d}{dt}\left(\frac{\partial\mathcal{L}}%
               {\partial(\frac{d}{dt}{\delta}R)}%
               \right) %
= 0
\label{EqA37}%
\end{equation}
leads, with the Lagangian of Eq.~(\ref{EqA36}), to the
Euler-Lagrange equation
\begin{equation}
  -c_{11}\frac{2{\delta}R}{R^2}
= {\rho_{2D}}\frac{d^2}{dt^2}{\delta}R \,.%
\label{EqA38}
\end{equation}
With the ansatz
\begin{equation}
  {\delta}R = {\delta}R_0~e^{i{\omega}t}%
\label{EqA39}
\end{equation}
we obtain
\begin{equation}
 {\omega} = \frac{1}{R} \sqrt{\frac{2c_{11}}{\rho_{2D}}} \,,%
\label{EqA40}
\end{equation}
which is identical to Eq.~(\ref{Eq7}).


\subsection{Radial breathing mode of carbon nanotubes}

Let us now consider a carbon nanotube that has been rolled up from
a rectangular graphene nanoribbon of width $2{\pi}R_0$, length
$L$, and the 2D mass density ${\rho_{2D}}$. According to continuum
elasticity theory, the total bending strain energy of any such
nanotube is given by~\cite{DT071}
\begin{equation}
U_b = \frac{\pi D L}{R_0} \,, %
\label{EqA41}
\end{equation}
where $D$ is the flexural rigidity of graphene. Increasing the
radius by the small amount ${\delta}R$, while still
${\delta}R/R_0<<1$, changes the potential energy of the nanotube
with respect to the value at $R_0$ by
\begin{equation}
{\Delta}U_b({\delta}R) = \frac{\pi D L}{R_0} %
              \left[-\frac{{\delta}R}{R_0}+%
              \left(\frac{{\delta}R}{R_0}\right)^2\right] \,, %
\label{EqA42}
\end{equation}
where higher than quadratic terms have been neglected in the
Taylor expansion.

We consider the initial nanotube of radius $R_0$ to be free of
tensile strain energy. In this case, the change in tensile strain
energy associated with the radius increase by ${\delta}R$ is
\begin{equation}
{\Delta}U_t({\delta}R) = \frac{1}{2} %
      \left[ 2 \pi R_0 c_{11}\right] L %
      \left(\frac{{\delta}R}{R_0}\right)^2 \,.%
\label{EqA43}
\end{equation}
The equilibrium value of ${\delta}R$ can be obtained by minimizing
${\Delta}U_b+{\Delta}U_t$,
\begin{equation}
\frac{\partial({\Delta}U_b+{\Delta}U_t)}{\partial({\delta}R)}=0\,.
\label{EqA44}
\end{equation}
This leads to
\begin{equation}
{\delta}R {\approx} \frac{D}{2c_{11}R_0}%
\label{EqA45}
\end{equation}
and the equilibrium value of the nanotube radius becomes
$R=R_0+{\delta}R$
\begin{equation}
R = R_0 \left(1+\frac{D}{2c_{11}R_0^2}\right) \,.%
\label{EqA46}
\end{equation}
$R$ is the radius, around which the RBM vibrations take place. We
use the total potential energy $U$ of the nanotube at the
equilibrium radius $R$ as the reference energy. Then, changing the
equilibrium nanotube radius $R$ by an arbitrarily small value
${\delta}R$ increases the potential energy of the nanotube by
\begin{equation}
U = \frac{1}{2} [2 \pi R L] \left(\frac{D}{R^4}+\frac{c_{11}}{R^2}\right) %
    ({\delta}R)^2 \,.%
\label{EqA47}
\end{equation}
The (radial) kinetic energy of a radially expanding or contracting
nanotube, shown schematically in the inset of Fig.~\ref{fig3}(b),
is given by
\begin{equation}
T = \frac{1}{2} \left[ 2 \pi R {\rho_{2D}} L\right] %
    \left( \frac{d}{dt}{\delta}R \right)^2 \,.
\label{EqA48}
\end{equation}
The Lagrangian for this motion is
\begin{eqnarray}
\label{EqA49}%
\mathcal{L}
&=& T - U \\
&=& [\pi R {\rho_{2D}} L] %
    \left(\frac{d}{dt}{\delta}R \right)^2
   - [\pi R L] \left(\frac{D}{R^4}+\frac{c_{11}}{R^2}\right) %
    ({\delta}R)^2 \,. \nonumber%
\end{eqnarray}
Using Hamilton's principle of Eq.~(\ref{EqA37}), we get the
Euler-Lagrange equation
\begin{equation}
  {\rho_{2D}} \frac{d^2({\delta}R)}{dt^2}
= -\left(\frac{D}{R^4}+\frac{c_{11}}{R^2}\right) {\delta}R \,.%
\label{EqA50}
\end{equation}
Using the ansatz
\begin{equation}
  {\delta}R = {\delta}R_0~e^{i{\omega}t}%
\label{EqA51}
\end{equation}
and referring to the nanotube diameter $d=2R$, we obtain
\begin{equation}
 {\omega} = \frac{2}{d} %
            \sqrt{ \frac{c_{11}}{\rho_{2D}} %
            \left( 1+\frac{D}{c_{11}R^2}\right)} \,.%
\label{EqA52}
\end{equation}
For a nanotube with a typical diameter $d=1$~nm, using the values
of $c_{11}$ and $D$ for graphene, we find that the bending
correction $4D/(c_{11} d^2){\approx}1{\times}10^{-3}$ is
negligibly small. In this case, we obtain
\begin{equation}
 {\omega} = \frac{2}{d} %
            \sqrt{ \frac{c_{11}}{\rho_{2D}} } \,,%
\label{EqA53}
\end{equation}
which is identical to Eq.~(\ref{Eq8}).

%
\section*{Acknowledgments}

We acknowledge useful discussions with Jie Guan and thank Garrett
B. King for assistance with the artwork in Fig.~\ref{fig1}. A.G.E.
acknowledges financial support by the South African National
Research Foundation. D.L. and D.T. acknowledge financial support
by the NSF/AFOSR EFRI 2-DARE grant number EFMA-1433459.


%

\end{document}